# Improving the Throughput of the AES Algorithm with Multicore Processors

Angelo Barnes, Ryan Fernando, Kasuni Mettananda and Roshan Ragel
Department of Computer Engineering
University of Peradeniya
Peradeniya 20400 Sri Lanka

*Abstract—* AES, Advanced Encryption Standard, can be considered the most widely used modern symmetric key encryption standard. To encrypt/decrypt a file using the AES algorithm, the file must undergo a set of complex computational steps. Therefore a software implementation of AES algorithm would be slow and consume large amount of time to complete. The immense increase of both stored and transferred data in the recent years had made this problem even more daunting when the need to encrypt/decrypt such data arises. As a solution to this problem, in this paper, we present an extensive study of enhancing the throughput of AES encryption algorithm by utilizing the state of the art multicore architectures.

We take a sequential program that implements the AES algorithm and convert the same to run on multicore architectures with minimum effort. We implement two different parallel programmes, one with the fork system call in Linux and the other with the pthreads, the POSIX standard for threads. Later, we ran both the versions of the parallel programs on different multicore architectures and compared and analysed the throughputs between the implementations and among different architectures. The pthreads implementation outperformed in all the experiments we conducted and the best throughput obtained is around 7Gbps on a 32-core processor (the largest number of cores we had) with the pthreads implementation.

## I. INTRODUCTION

With the development of the multicore processors, running applications in multiple instances with the objective of improving their performance (that is throughput) has become common. As the most widely used encryption algorithm and given that it is regularly used in *Internet* based applications the performance of AES matters a lot in places where data is stored and transmitted in the encrypted form. In this paper we explore parallelizing a software implementation of AES thereby increasing the throughput by utilizing the available cores on multicore processors.

If the main objective of our project is to improve the throughput of an AES implementation with the help of multicore processors, our secondary objective is to evaluate two different parallel implementations (fork system call and POSIX threads) through comparing their performance. We are comparing the results obtained by running the AES algorithm in parallel with different number of processes using machines with different number of cores and evaluate the efficiency of using multiple cores for the AES algorithm. The idea here is to see whether the throughput can be improved by processing the encryption, simultaneously as parallel instances. For this we use several CPUs to manipulate the multi-threading ability. After measuring the execution time of different file sizes on different processors, we compare the results. As our main objective is to find whether the parallelized implementation is efficient, we need to run our parallelized AES implementation on different processors. Therefore we select a number of different CPU configurations according to the availability and the configurations are a dual core processor, a quad core processor and a 32-core processor. To achieve our secondary objective, we use both fork and pthreads implementations in our experiments and compared the throughput variations.

The rest of the paper is organized as follows: Section II presents related work and Section III presents the background. Design and implementation are presented in Section IV and how the experiments are performed is presented in Section V. Section VI presents the results and Section VII presents the discussion of the results. In Section VIII we conclude the paper.

## II. RELATED WORK

Most of the work done in accelerating AES is focused on hardware implementation in FPGAs and using graphic processing units (GPGPU) although there are a few CPU based acceleration techniques available.

Kotturi, Yoo and Blizzard [1] have presented a hardware-efficient design, increasing the throughput for the AES algorithm using a high-speed parallel-pipelined architecture. By using an efficient inter-round and intra-round pipeline design, their implementation has achieved a high throughput of 29.77 Gbps in encryption.

Huang, Chang, Lin and Tai [2] have proposed a 32-bit AES implementation in small Xilinx FPGA Chip (Spartan-3 XC3S200). It uses 148 slices, 11 Block RAMs (BRAMs) and achieves a throughput of 647 Mbps at 278 MHz working frequency. Also they have presented a 128-bit AES implementation in FPGA (Virtex-II XC2VP20) by parallel operations of four above 32-bit AES is also presented. In [3] Gielata, Russek and Wiatr have investigated hardware implementation of AES-128 cipher standard on FPGA technology. The investigations involved simulations and synthesis of VHDL code utilizing Virtex4 series of Xilinx. In [4] Mali, Novak and Biasizzo have done a hardware implementation of the AES algorithm developed for an external data storage unit in a dependable application. The AES algorithm was implemented in FPGA using the development board Celoxica RC1000 and development suite Celoxica DK. Christopher and Anantha [7] also have used FPGA to enhance the encryption of AES encryption. They have developed a freeware, high-throughput, parallel implementation of the AES algorithm for resource-limited hardware. They have compared the throughput results to advertised industry data throughput rates for comparable hardware settings.

Manavski [9] has used an NVIDIA GeForce 8800 GTX to run AES using the CUDA platform and also have used an Intel Pentium IV 3.0 GHz processor to run a sequential implementation of AES as a reference. The final result was that the GPU implementation out-performed the CPU implementation and the maximum throughput was obtained when the input file size was the largest. However, Manavski has only compared a sequential implementation running on a CPU and a parallelized implementation using GPGPU. We are concentrating on running a parallelized implementation of AES using the multicore feature of new CPUs. The results are expected to be different between a single core sequential implementation and a multicore parallelized implementation.

Bernstein and Schwabe [5] have presented new speed records for AES software (10.5 cycles/byte), taking advantage of architecture-dependent reduction of instructions used to compute AES and microarchitecture-dependent reduction of cycles used for those instructions. Osvik, Bos, Stefan and Canright [6] have presented new software speed records for AES-128 encryption for architectures at both ends of the performance spectrum (microcontrollers vs. GPGPUs).

Lee and Chen have worked on achieving the fastest software implementation of AES using a general-purpose processor. They have focused on improving the parallelism and performance of the software implementation. They are proposing an enhanced parallel table lookup instruction (Pread) that would achieve the fastest software encryption and decryption of AES [8].

In our research we are moving further with the existing work and trying out a parallelized AES implementation on multicore processors utilizing their inherent parallelism available and hence improving the throughput of the encryption. We take a sequential implementation of the AES algorithm and convert it into a multicore version with minimum effort. The multicore version utilises the existing parallel thread libraries for implementation and therefore the effort is minimal. We further go on to evaluate and compare the performance of two different such libraries (fork system call vs. POSIX standard for threads). Therefore, we try to improve the throughput of the AES algorithm with a software implementation with minimum changes to an existing sequential AES program.

### III. BACKGROUND

#### A. AES (Advanced Encryption Standard)

In the AES encryption process, a number of repetitions are used in transformation rounds. Each round consists of several processing steps. Below is the high level description of the AES algorithm and the steps are the following as shown in Fig. 1 [10].

1) Key Expansion
2) Initial Round
    I. add round key
3) Rounds
    I. sub bytes
    II. shift rows
    III. mix columns
    III. add round key
4) Final Round

I. sub bytes
II. shift rows
III. add round key

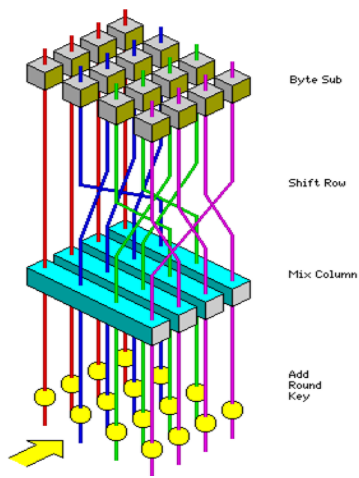

Figure 1.  Steps of AES process (taken from [10])

### B. Multithreading

Multithreading allows multiple threads to exist within the context of a single process. These threads share the process' resources but are able to execute independently. This advantage of a multithreaded program allows it to operate faster on computer systems that have multiple CPUs, CPUs with multiple cores, or across a cluster of machines because the threads of the program naturally lend themselves to truly concurrent execution.

## IV. DESIGN AND IMPLEMENTATION

### A. Overview

We chose an existing AES software implementation, which was implemented in C language and extended it to support and utilize multicore platform.

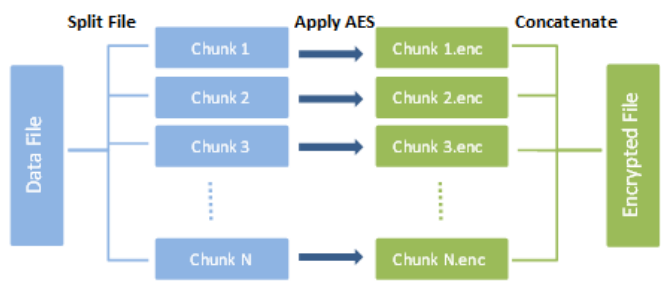

Figure 2.  Overview of the Methodology

The AES implementation is further developed and modularized (as shown in Fig. 2) so that it can be able to take the number of cores as the input, divide the data into smaller chunks according to that number and perform the encryption in parallel, so that the overall throughput is much larger compared to its single instance implementation. Then we used three different CPUs, which consist of different number of cores and tested the implemented algorithm with variable input data. Tests were conducted with variable input data sizes. Comparison of performances among the CPUs was conducted. Fig. 3 gives an overview of the methodology, which we are following in this phase. The name of the file to be encrypted, the number of threads, and the number of times we need to test are the inputs to the system.

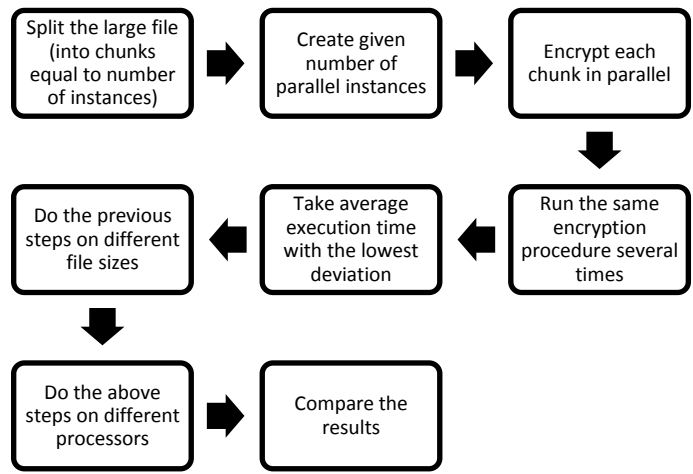

Figure 3.  Overview of the Methodology

As shown in Fig. 3, the system will split the large file into chunks equal to the number of parallel instances to be run. Then the encryption is done in parallel with the given number of processes or threads. Then this multithreaded encryption process is repeated a number of times and the average execution time is calculated by eliminating results that has higher deviation to make sure that the measurements recorded are fair.

### B. Parallel Design and Implementation of AES

In C language there are two ways of creating parallel programmes: one, using the *fork()* system call and two, using *pthread_create()* (using POSIX thread library). The main difference is, when using pthread, all threads within a process share the same address space, while fork returns two completely independent copies of the original process which have their own address spaces, with its own copies of its variables, which are completely independent of the same variables in the other process. Under a Unix-like system, *fork()* is a system call when used will create a child process which is alike the parent process. On the other hand, pthreads are defined as a set of C language programming types and procedure calls. Compared to the cost of creating and managing a process using *fork()*, a thread can be created with much less operating system overhead. Therefore, managing threads requires fewer system resources than managing processes from *fork()*.

Therefore when developing our AES parallel program, we used both the methods in two different instances and finally came up with two different implementations, so that we can test both and decide which one is the best. In the pthread implementation after breaking the file into several parts, threads were created using *pthread_create()* according to the number of parts and each part was encrypted in the newly created thread. After the threads finished their work they were joined using *pthread_join()* and the encrypted files were joined together to form a single file. In fork implementation the command *fork()* was called according to the number of threads. In a loop *fork()* was called multiple times and in each child process the main encryption function was run and each split chunk of the file was encrypted in parallel. After creation of all the child processes the system waits for each process to complete by calling the method *waitpid()* for each process ID in a loop, where a new process ID is created at each call of successful *fork()*. After completion of all the processes the newly created encrypted files were concatenated, and the final encrypted file was obtained.

## V. EXPERIMENTS

As we have implemented the parallel AES code in two different ways (using fork and pthread) we tested both the implementations and obtained the readings. The process was done as described earlier using Fig. 2. The same process was completed with various file sizes and the average execution time is measured for each process. Then the whole process was repeated in three different processors with the following configurations:

1. Intel Core 2 Duo processor (with 2 single threaded cores)
2. Intel Core i3 (with 2 cores, supporting 4 threads)
3. 4 x Intel® Xeon® X7560 (with 32 cores, supporting 64 threads)

We ran the largest numbers of instances (up to 32) only on the 32-core processer as we know that the other processors (with 2-cores) cannot run that many numbers of threads simultaneously. Finally, all the results we obtained were compared and analysed.

When we get the average execution time, we considered only the reasonably close (for recording fair measurements) execution times. We ran the same process (with same file and same number of threads) for a given number of times. Then by looking at all the time measurements we eliminated the ones that had larger deviations. We used the rest of the time measurements to calculate the average time taken to encrypt that file for a given number of threads and finally the throughput values.

## VI. RESULTS

In this section, the results we obtained from the testing in three different processors with both different implementation methods we chose are reported. As mentioned earlier, the measurements are taken for different file sizes so that we can evaluate the scalability of our approach.

### A. Core 2 Duo Processor (2 cores, 2 threads)

Fig. 4 depicts the throughput variation against the data size in a dual core processor for the fork implementation and Fig. 5 depicts the same variation for pthread implementation.

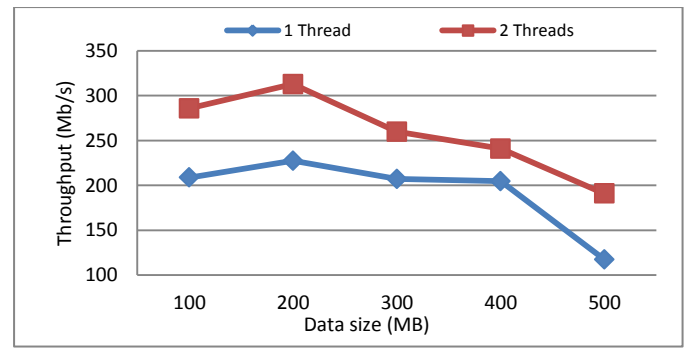

Figure 4. Throughput vs. Datasize for Core 2 Duo Processor using fork

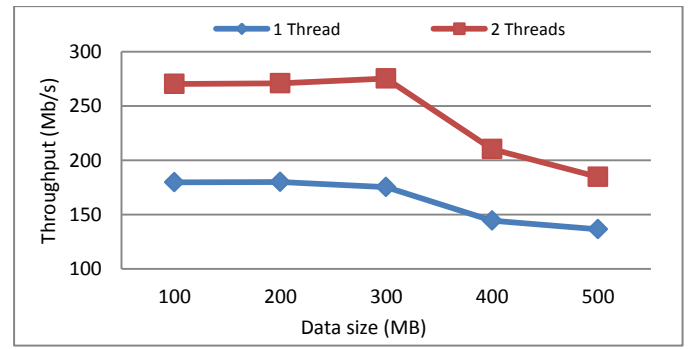

Figure 5. Throughput vs. Datasize for Core 2 Duo Processor using pthread

### B. Core i3 ( 2 cores, 4 threads)

Fig. 6 depicts the throughput variation against the data size for in an *Intel Core i3* processor for the fork implementation and Fig. 7 depicts the same variation for the pthread implementation.

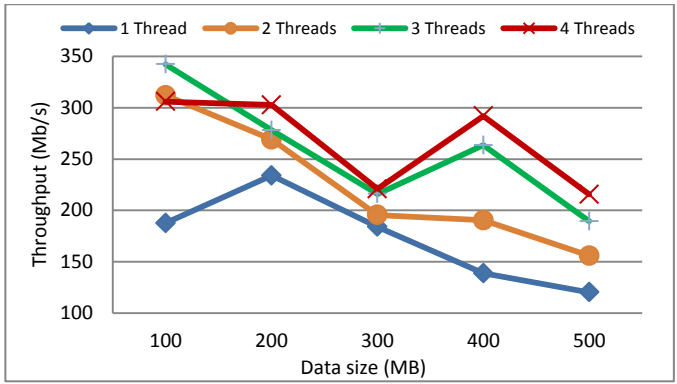

Figure 6.  Throughput vs. Datasize for Core i3 Processor using fork

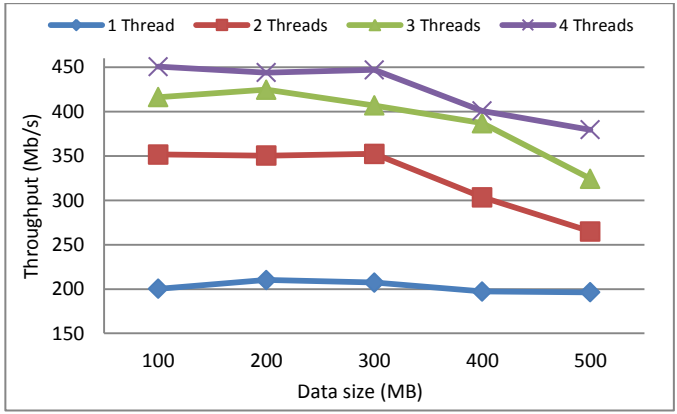

Figure 7.  Throughput vs. Datasize for Core i3 Processor using pthread

## C. 4 x Intel® Xeon® X7560 Processors (32 cores, 64 threads)

Fig. 8 depicts the throughput variation against the data size for a 32-core machine for the fork implementation and Fig. 9 depicts the same variation for pthread implementation.

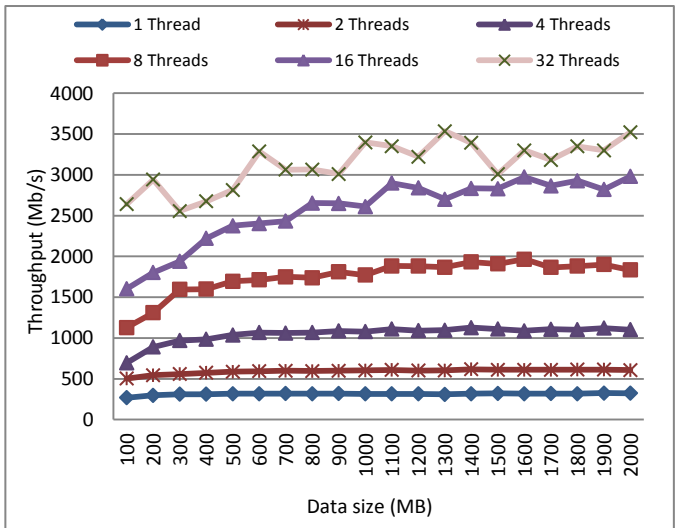

Figure 8.  Throughput vs. Datasize for 32 Core processor using fork

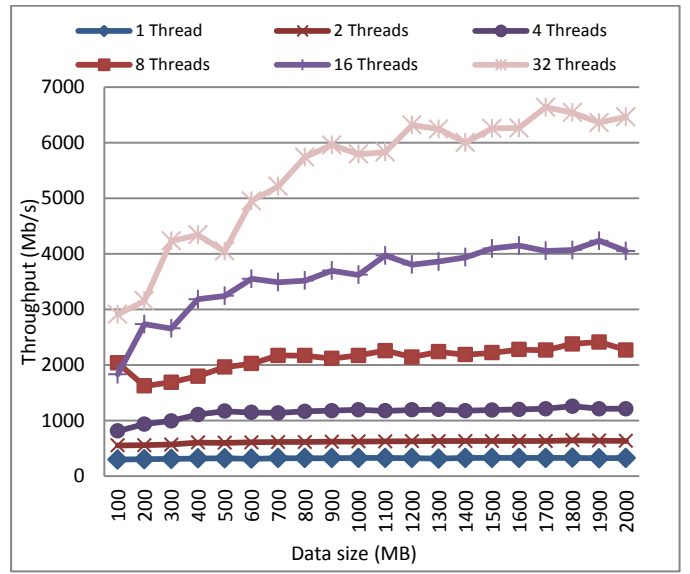

Figure 9.  Throughput vs. Datasize for 32 Core processor using pthread

We summarize the best throughputs (column 3) of each processor (per method: *fork* and *pthread*) of all the experiments in Table I. In addition, we have also calculated the throughput per core and reported in column 4 of the table.

TABLE I.  RESULTS SUMMARY

|  | Processor | Best throughput (Mb/s) | Through/core (Mb/s per core) |
| --- | --- | --- | --- |
| *fork* | *Core 2 Duo* | 312 | 156 |
|  | *Core i3* | 342 | 171 |
|  | *32-core* | 3535 | 110 |
| *pthread* | *Core 2 Duo* | 276 | 138 |
|  | *Core i3* | 450 | 225 |
|  | *32-core* | 6637 | 207 |

## VII. DISCUSSION OF THE RESULTS

From the results we obtained, we can clearly see that parallelizing the algorithm can increase the throughput. From each graph we presented, we observe the followings:

- In both fork and pthread implementations the throughput has been increased with the number of instances.
- In *Core 2 Duo* processor the throughput increases for 2 threads.
- In *Core i3* processor the throughput increases for 4 threads.
- In 32-core machine the throughput increases with the number of threads for 32 threads.
- Smaller file sizes gave more throughputs in lesser number of threads; however, when we used large number of threads it is the opposite. The throughput increased with the file size and higher number of execution instances.

- When testing with *Core 2 Duo* and *Core i3* the throughput achieved decreased whilst the size of the file increases.
- But when testing with the 32-core CPU, greater file sizes gained greater throughputs.
- *Pthread* implementation is 1.90 ($\approx 2$) times better than fork implementation when using 32 threads with 2GB file size.
- However, fork showed a slight throughput increase than *pthread* when testing with the Core 2 Duo CPU (showed a 36 Mb/s throughput increase).

By comparing all the results we obtained and from the graphical representations presented above and the summary table, the inferences we can come up with are:

- Both the implementations increase the throughput by parallelizing with multiple instances.
- Generally the pthread implementation gives a higher throughput than the fork implementation and therefore pthread almost always performs better than fork.
- As *Core 2 Duo* and *Core i3* processors can only run 2 threads and 4 threads respectively, the throughput is not increased after this number.
- Generally, the throughput increases with file size (with the higher number of instances).
- The highest throughput we gained was 6637 Mb/s. This was obtained by running 32 threads on the 32-core machine using the pthread implementation.

## VIII. CONCLUSION AND FUTURE WORK

This project was performed in order to check whether we can enhance the performance of AES encryption by parallelizing it and run them on multi-threaded processors. Therefore, after all the experiments conducted and results analysed, we can finally come to the following conclusions: (1) by parallelizing the AES algorithm the throughput can be considerably enhanced and performance also can be increased; and (2) out of fork and pthread implementations, pthread generally outperforms the fork implementation. As for the future work we are expecting to run the parallelized AES algorithm on an NVIDIA GPU and examine the throughput and performance enhancements. Then we will compare those results over the results we obtained from the CPUs.